\documentclass{iau} 
\usepackage{graphicx}
\usepackage{natbib}
\usepackage{amsmath,amssymb}
\usepackage{gensymb}
\usepackage{chngcntr}
\renewcommand{\theequation}{\arabic{equation}}
\counterwithout*{equation}{section}

\def\apj{\textit{ApJ}}
\def\apjl{\textit{ApJ} (Letters)}
\def\mnras{\textit{MNRAS}}
\def\apjs{\textit{ApJS}}
\def\nat{\textit{Nature}}
\def\aj{\textit{AJ}}

\def\aap{\textit{A$\mathit{\&}$A}}

\def\araa{\textit{ARAA}}

\def\procspie{\textit{Proc. SPIE}}

\def\frass{\textit{Front. Astron. Space Sci.}}
\def\natas{\textit{Nature Astronomy}}

\newcommand{\oiii}{[O\,\textsc{iii}]}

\newcommand{\cii}{[C\,\textsc{ii}]}

\newcommand{\ha}{H$\alpha$}

\title[GV 20.~~AGN outflows and feedback from high to low z] 
{The physical properties and impact of AGN outflows from high to low redshift\thanks{Based on observations made with ESO Telescopes at the La Silla Paranal Observatory under program ID 094.B-0321(A).}
}

\author[Giacomo Venturi et al.]   
{Giacomo Venturi$^{1,2}$
  \and Alessandro Marconi$^{3,2}$
 }

\affiliation{$^1$Instituto de Astrof{\'{i}}sica, Pontificia Universidad Cat{\'{o}}lica de Chile, Avda. Vicu{\~{n}}a Mackenna 4860, 8970117, Macul, Santiago, Chile \\ email: {\tt gventuri@astro.puc.cl} \\[\affilskip]
$^2$INAF - Osservatorio Astrofisico di Arcetri, Largo E. Fermi 5, I-50125, Firenze, Italy 
 \\[\affilskip]
$^3$Dipartimento di Fisica e Astronomia, Universit{\`{a}} degli Studi di Firenze, Via G. Sansone 1, I-50019, Sesto Fiorentino, Firenze, Italy \\[\affilskip]
}

\pubyear{2020}
\volume{359}  
\jname{Galaxy evolution and feedback across different environments}
\editors{T. Storchi-Bergmann, R. Overzier, W. Forman \& R. Riffel, eds.}
\begin{document}

\maketitle

\begin{abstract}
Feedback from active galactic nuclei (AGN) on their host galaxies, in the form of gas outflows capable of quenching star formation, is considered a major player in galaxy evolution. However, clear observational evidence of such major impact is still missing; uncertainties in measuring outflow properties might be partly responsible because of their critical role in comparisons with models and in constraining the impact of outflows on galaxies. Here we briefly review the challenges in measuring outflow physical properties and present an overview of  outflow studies from high to low redshift. Finally, we present highlights from our MAGNUM survey of nearby AGN with VLT/MUSE, where the  high intrinsic spatial resolution (down to $\sim$10 pc) allows us to accurately measure the physical and kinematic properties of ionised gas outflows.
\end{abstract}

\firstsection 
\section{Introduction}
Feedback from active galactic nuclei (AGN) on their host galaxies is considered a critical element to explain many key observed properties of galaxies during  their  evolution  over cosmic time.
These are: 1) the observed discrepancy in the luminosity function of galaxies at high masses between  observations and predictions from models without AGN feedback  (e.g. \citealt{Kormendy:2013aa} and references therein); 2) the scaling relations observed between the mass of supermassive black holes (BHs) and the properties of their host galaxies (e.g. \citealt{Behroozi:2019aa}); 3) the similarity between the BH accretion history and the star formation (SF) history throughout cosmic time (e.g. \citealt{Aird:2015aa}); 4) the bimodality of galaxies, divided in two distinct populations in the colour versus stellar mass diagram, i.e. the so-called ``blue cloud'' and the ``red sequence'' (e.g. \citealt{Schawinski:2014aa}).
Feedback from AGN is thus routinely included in models and simulations of galaxy formation and evolution, as it is able to explain the above properties (e.g. \citealt{Ciotti:2010aa}, \citealt{Schaye:2015aa}).
AGN feedback is commonly divided in  two different modes (see e.g. \citealt{Fabian:2012aa} and references therein): a ``radiative'' (or ``quasar'') mode, operating during a luminous AGN phase through powerful outflows which sweep away the gas reservoir from the host galaxy, thus quenching SF; a ``kinetic'' (or ``radio'') mode, acting steadily on longer timescales through radio jets which heat the gas halo surrounding massive galaxies, thereby preventing its cooling and reaccretion on the host and consequently further SF. While clear evidence for the kinetic mode has been found in massive central cluster galaxies, where cavities in the X-ray emitting hot ionised gas, filled by radio jets propagating from the AGN, are observed (e.g. \citealt{McNamara:2000aa}, \citealt{Birzan:2012aa}), the radiative mode is more elusive and convincing evidence for the impact of outflows on host galaxies is still missing.

Here we provide a brief review on AGN outflows and their role in the context of AGN feedback from an observational point of view, focusing on  outflows on galactic scales, where their effects on  host galaxies are expected to be observable.

\section{Measuring outflow physical properties}\label{sec:out_props}
Measuring the physical properties of outflows with high accuracy and well established methods is of primary importance for constraining their  impact on host galaxies. Different assumptions in calculating outflow properties lead in fact to a large spread in their inferred kinematics and energetics. The spread remains quite large even when using the same method, since some outflow properties (such as density, extension, inclination) are often unknown and must be assumed (e.g. \citealt{Harrison:2018aa}). Such uncertainties and different adopted methods make then difficult the comparison between different works and with the predictions from models and simulations of outflows and feedback.

The mass outflow rate through a spherical (or conical or multi-conical) surface can be calculated from the fluid continuity equation, as:
\begin{equation}
    \dot{M} = \Omega r^2 \rho v \simeq f \dfrac{M v}{r}, \label{eq:moutrate}
\end{equation}
where $\Omega$ is the angle subtended by the outflow as seen by the  AGN, $r$ its radius, $\rho$ the outflowing gas density, $v$ its velocity, $M$ its mass and
 $f$ a factor depending on the outflow geometry, equal to 3 in the case of a volume filled with outflowing clouds or 1 for a simple shell  geometry (see e.g. \citealt{Maiolino:2012aa}).
 Velocity, mass and size of the outflow are thus what needs to be measured to determine the mass outflow rate.
 
 Velocities can be obtained through a multi-Gaussian fitting of the emission-line profiles to isolate the outflowing component. However, in some cases this approach may fail in isolating outflows;  parametric velocities from the total line profile can be used such as percentile velocities (e.g. $v_{10}$, the velocity containing 10$\%$ of the total emission line flux) and the corresponding velocity widths (W80 = $v_{90} - v_{10}$; e.g. \citealt{Harrison:2014aa}). Such measurements are based on the observed line-of-sight velocity distribution of the gas and thus, in order to try accounting for projection effects, the far wing of the line profiles (e.g. $v_{05}$) or a combination of this with the line width are often used.
 However, to properly recover the intrinsic velocities across the outflow, which also likely vary with distance and angle, a full kinematic modelling would be required (e.g. \citealt{Crenshaw:2015aa}).
 
 Measuring the outflow mass firstly deals with the challenge of isolating the fraction of spectral line profiles associated with the outflow, which may not be a trivial task, being affected by degeneracies. 
 The mass, in the case of molecular outflows, is mainly obtained from CO millimetre/sub-millimetre emission lines as $M_\mathrm{mol} = \alpha_\mathrm{CO} L_\mathrm{CO(1-0)}$, where the CO-to-H$_2$ conversion factor $\alpha_\mathrm{CO}$, usually assumed, is uncertain [$\sim$0.8$-$4 M$_\odot$ (K km s$^{-1}$ pc$^2$)$^{-1}$, based on values for local spiral galaxies or mergers; e.g. \citealt{Bolatto:2013aa}]; in the case of higher $J$ rotational transitions, uncertainties up to a factor of $\sim$3 are introduced by the conversion of their luminosity to $L_\mathrm{CO(1-0)}$  (e.g. \citealt{Brusa:2018aa}). H$_2$ near-IR emission lines only trace a minor warm transitory phase and are not representative of the total molecular mass budget (e.g. \citealt{Tadhunter:2014aa}). 
 The warm ionised outflow mass is usually obtained from the luminosity of optical emission lines, such as \oiii$\lambda$5007 or Balmer lines, and depends on electron temperature, element abundance and ionisation state (except for Balmer lines) and, above all, electron density (e.g. \citealt{Carniani:2015aa}), the latter constituting the largest source of uncertainty in the ionised mass determination (e.g. \citealt{Harrison:2018aa}). 
 
 Finally, the ability to measure the size of the outflow is strictly dependent on the spatial resolution of observations: in many cases outflows are often only barely resolved or completely unresolved and consequently the size must be assumed.
 
 \section{Outflows from high to low redshift}
 Pieces of evidence for the presence of outflows in AGN are found all the way from high to low $z$. Here we highlight some significant examples from the many studies available in the literature.
 
 {\underline{\it High redshift}}. 
 At $z$\,$\sim$\,6 very few indications of outflows exist so far. 
 Direct detection of outflows at high $z$ comes in the form of faint broad \cii\ wings in quasar (QSO) sub-mm spectra, both in few individual objects (e.g. \citealt{Maiolino:2012aa}) or stacked spectra (\citealt{Bischetti:2019aa}) or from OH in absorption in the far-IR (\citealt{Herrera-Camus:2020aa}).
 The cold gas kinematics of QSOs, traced by CO or \cii, indicates turbulent thick gas discs, possibly due to dynamically hot discs and/or outflows (\citealt{Pensabene:2020aa}).
 High-resolution ($\sim$400 pc) kinematics of a $z$\,$\sim$\,6.5 QSO revealed \cii\ cavities around the galaxy centre, a potential signature of QSO feedback in action (\citealt{Venemans:2019aa}).

 {\underline{\it ``Intermediate'' redshift}}.
 Outflows at $z$\,$\sim$\,1$-$3, around the peak of AGN and SF activity (``cosmic noon''), are mainly observed in the near-IR band, where the optical emission lines are redshifted. Evidence for outflows suppressing SF in the host was also found in a few objects, from the spatial anti-correlation between the distribution of \oiii\ and CO high velocity emission tracing outflows and narrow \ha\ emission  tracing SF (\citealt{Cresci:2015ab}, \citealt{Carniani:2016aa}, \citealt{Brusa:2018aa}), which has recently been questioned in one case (\citealt{Scholtz:2020aa}).
 A number of observational campaigns such as the WISSH survey of hyper-luminous QSOs (\citealt{Bischetti:2017aa}), the SUPER survey (\citealt{Circosta:2018aa}), the KASH-z survey of X-ray selected AGN (\citealt{Harrison:2016aa}), found ubiquitous ionised outflows at these redshifts. Moreover, the SINS zC-SINF plus KMOS3D (\citealt{Forster-Schreiber:2018aa,Forster-Schreiber:2019aa}) and the MOSDEF (\citealt{Leung:2019aa}) surveys found that the incidence of fast ionised outflows increases with stellar mass jointly with AGN fraction, and so their incidence in AGN is independent on stellar mass. 
 
 {\underline{\it Low redshift}}.
 A plethora of outflow studies exists at $z$\,$<$\,1. These span from QSOs around $z$\,$\sim$\,0.5 (e.g. \citealt{Villar-Martin:2011aa}) or $z$\,$\sim$\,0.2 (e.g. \citealt{Harrison:2014aa}) down to the very local Universe.
 In particular, recent spatially resolved studies of local AGN, mostly making use of integral field spectroscopic observations, proved themselves crucial in characterising and constraining the properties of ionised outflows, by providing the extension and geometry of outflows and allowing to resolve their kinematics.
 This has been possible both on small scales, through few-arcsec scale integral field spectrographs (such as Gemini GMOS and NIFS; e.g. \citealt{Riffel:2015aa}, \citealt{Freitas:2018aa}) or {\it HST} imaging + spectroscopy (e.g. \citealt{Revalski:2018aa}), and on large scales where, in particular, the optical and near-IR integral field spectrograph MUSE at VLT has been a breakthrough. In fact, its combination of wide field of view (1$'\times$1$'$) and  spectral coverage provides a wealth of gas diagnostic lines which, together with the 8m-telescope high sensitivity, allows to study in  details the outflow kinematics  and the gas physical properties from tens of pc up to few kpc, as shown for instance by our MAGNUM survey (e.g. \citealt{Venturi:2017aa,Venturi:2018aa}, \citealt{Mingozzi:2019aa}) - which we will delve into in Section \ref{sec:MAGNUM} - and by  the AMUSING++ compilation (\citealt{Lopez-Coba:2020aa}).
 
 
Neutral outflows in local objects are either studied in their neutral atomic phase through the Na\,\textsc{i}\,D optical absorption line doublet (e.g. \citealt{Perna:2019aa}) and H\,\textsc{i} (\citealt{Morganti:2016aa}), or, mostly, in their molecular phase through broad wings of species like CO or OH, tracing the cold gas reservoir (e.g. \citealt{Cicone:2014aa}, \citealt{Garcia-Burillo:2014aa}, \citealt{Lutz:2020aa}), and H$_2$ for what concerns the warm molecular phase (e.g. \citealt{Emonts:2014aa}).
 The origin of molecular gas in outflows, with supersonic velocities of $\sim$1000 km/s and mass outflow rates of a few 1000s M$_\odot$/yr, and whether it is accelerated directly at the source or formed by gas cooling within the outflow, is debated. Some models (e.g. \citealt{Zubovas:2014aa}) predict that outflowing hot gas is unstable and would eventually lead to a two-phase medium, formed by cold dense molecular clumps surrounded by hot tenuous gas. Moreover, efficient gas cooling would result in stars forming within the outflow itself, with potential important implications for galaxy formation and evolution. Such phenomenon was recently confirmed  from observations of  gas ionised by young stars within outflows, or through direct kinematic signatures of outflowing  young stars (e.g. \citealt{Maiolino:2017aa}, \citealt{Gallagher:2019aa}).
 
{\underline{\it Multi-phase studies}}. Having information from all the gas phases is of primary importance for determining the total mass and energetic budget of the outflows, comparing with predictions from feedback models and determining their impact on host galaxies. However, very few multi-phase outflow studies exist up to now (\citealt{Cicone:2018aa}). The prototypical example of a multi-phase outflow is Mrk 231, the closest QSO known, studied in the ionised phase (\citealt{Rupke:2011aa}), in the neutral atomic phase from both Na\,\textsc{i}\,D (\citealt{Rupke:2011aa}) and H\,\textsc{i} (\citealt{Morganti:2016aa}) and in the molecular phase from both CO (\citealt{Feruglio:2015aa}) and OH (\citealt{Fischer:2010aa}).
 When the total mass budget of the galactic-scale outflow is available, it is indeed possible, by comparing with the nuclear unresolved X-ray outflow, to infer whether the outflow is, for instance, momentum- or energy-conserving (e.g. \citealt{Feruglio:2015aa}).
 
 {\underline{\it Summary of local outflow properties}}.
 Studies of $z$\,$<$\,1 AGN from SDSS (e.g. \citealt{Mullaney:2013aa}, \citealt{Woo:2016aa}) and MaNGA (\citealt{Wylezalek:2020aa}) have shown that ionised outflows are ubiquitous, that their fraction is higher in AGN than in inactive galaxies and that it increases with AGN luminosity and/or Eddington ratio. Despite their widespread presence, their global effect on SFR in the host does not seem to be significant (e.g. \citealt{Balmaverde:2016aa}), except maybe in high-SFR objects (\citealt{Wylezalek:2016aa}).
 
 Studies of  samples of AGN outflows in multiple gas phases (\citealt{Carniani:2015aa}, \citealt{Fiore:2017aa}, \citealt{Fluetsch:2019aa}) found relations for outflow velocity, mass, momentum and kinetic energy rates with AGN luminosity across several orders of magnitude and in all phases analysed. Moreover, they found that ionised outflow masses are usually $\sim$10 times smaller than those of molecular outflows in AGN, while they are similar in SF galaxies. The mass in the neutral atomic phase is generally comparable with that in the molecular phase. They also concluded that, while gas depletion times increases with AGN luminosity, the outflow escape fraction is small ($\lesssim$\,5$\%$) and so most of the outflowing gas is not able to leave the galaxy.
 Finally, short flow times ($\sim$10$^6$~yr) suggest intermittent AGN activity and, indeed, evidence for ``fossil'' outflows (i.e. from a faded AGN stronger in the past) is found (e.g. \citealt{Fluetsch:2019aa}, \citealt{Audibert:2019aa}).

 \section{MAGNUM survey}\label{sec:MAGNUM}

Here we present some highlights from our MAGNUM survey (Measuring Active Galactic Nuclei Under MUSE Microscope; e.g. \citealt{Venturi:2017aa}, \citealt{Mingozzi:2019aa}) of nearby Seyfert galaxies observed with the wide-field optical and near-IR integral field spectrograph Multi Unit Spectroscopic Explorer (MUSE; \citealt{Bacon:2010aa}) at the Very Large Telescope (VLT). The vicinity of the targets, combined with the unique capabilities of MUSE, allowed us to map the ionised gas down to $\sim$10 pc in several nebular emission lines, revealing ubiquitous kpc-scale outflows, and to study their physical properties.


\begin{figure}
\begin{center}
 \includegraphics[width=5.2in, trim={0 8cm 0 0},clip]{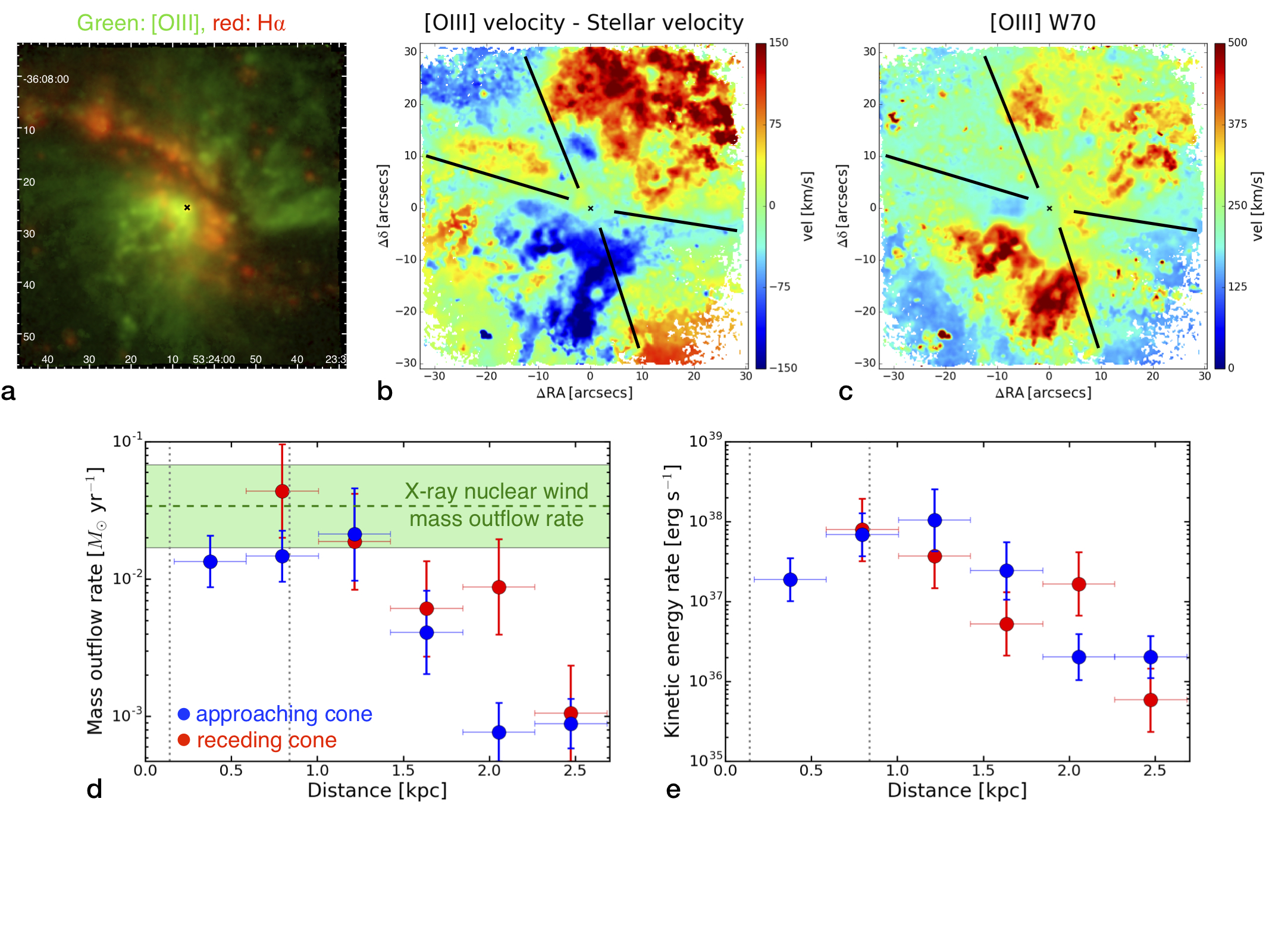} 
 \caption{VLT/MUSE maps of NGC 1365. a) Emission-line maps of ionised gas, \oiii$\lambda$5007 (green) and \ha\ (red). b) \oiii\ velocity, after subtraction, spaxel-by-spaxel, of the stellar velocity. c) \oiii\ W70 (line width). The black lines are meant to guide the eye. d) Radial profile of the ionised mass outflow rate and e) kinetic energy rate as a function of distance from the AGN. The green dashed line indicates the mass outflow rate of the nuclear unresolved X-ray wind.}
   \label{fig:fig1}
\end{center}
\end{figure}

{\underline{\it Dissecting the properties of the ionised outflow in NGC 1365}}. In \cite{Venturi:2018aa} we focused on the detailed study of a specific source, NGC 1365, a barred Seyfert galaxy located at 17.3 Mpc from Earth hosting a low-luminosity AGN ($\sim$2$\times$10$^{43}$ erg/s). MUSE observations, covering its $\sim$5.3$\times$5.3~kpc$^2$ central region, are shown in Fig.~\ref{fig:fig1}. Panel a) reports the \oiii$\lambda5007$ (green) and \ha\ (red) emission, the former mostly tracing the gas ionised by the AGN, the latter stemming from SF regions. The ionisation cones host a bi-conical ionised outflow, as shown in panels b) and c), where the maps of \oiii\ velocity and of \oiii\ W70, measure of the line velocity width (difference between the 85$\%$ and 15$\%$ percentile velocities of the total line profile), are displayed, respectively. 
In order to isolate the gas motions in excess to rotation, the stellar velocity has been subtracted spaxel-by-spaxel from the \oiii\ velocity.
We then extracted radial profiles of the outflow kinematics and energetics as a function of distance from the AGN. We show the mass outflow rate and the kinetic energy rate in panels d) and e).
The mass outflow rate was calculated by radially slicing the two outflowing cones and using Eq. \ref{eq:moutrate}, where f$=$1, $r$=$\Delta$r is the shell radial width and $M$ and $v$ the outflow mass and velocity in each radial slice, respectively, obtained from the centroid and \ha\ flux of the fitted Gaussian component associated to the outflow (blueshifted to the south-east, redshifted to the north-west), respectively.
The comparison of the galactic-scale outflow with the mass outflow rate of the highly-ionised nuclear unresolved outflow, measured from Fe\,\textsc{xxv} and  Fe\,\textsc{xxvi} X-ray absorption lines and having velocities of $\sim$3000 km\,s$^{-1}$, provides insights on the outflow driving mechanisms. We could thereby exclude a momentum- or an energy-driven mechanism as the origin of the outflow acceleration.
The former is in fact predicted on scales $\lesssim$\,1~kpc (e.g. \citealt{King:2015aa}), smaller than those observed for the outflow in NGC 1365. The latter is ruled out since energy conservation between the large-scale kinetic rate in all gas phases and the nuclear one would require a fraction of neutral atomic and molecular gas a factor $\gtrsim$10$^3$ larger than that in the ionised phase. On the other hand, direct acceleration by the AGN radiation pressure on dusty clouds could, in principle, be a possible driver of the outflow. The AGN photon momentum ($L_\mathrm{bol}/c$) is in fact $\sim$20 times larger than the peak of the optical galactic outflow momentum rate ($\dot{p}$), and models predict $\dot{p} \! \sim$1$-$5$ L_\mathrm{bol}/c$ (e.g. \citealt{Ishibashi:2018aa}), implying that a fraction of neutral atomic plus molecular gas a factor 20$-$100 larger than the ionised one would be necessary, which is feasible based on literature studies (e.g. \citealt{Fluetsch:2019aa}).


\begin{figure}
\begin{center}
 \includegraphics[width=5.3in, trim={0 3cm 0 3cm},clip]{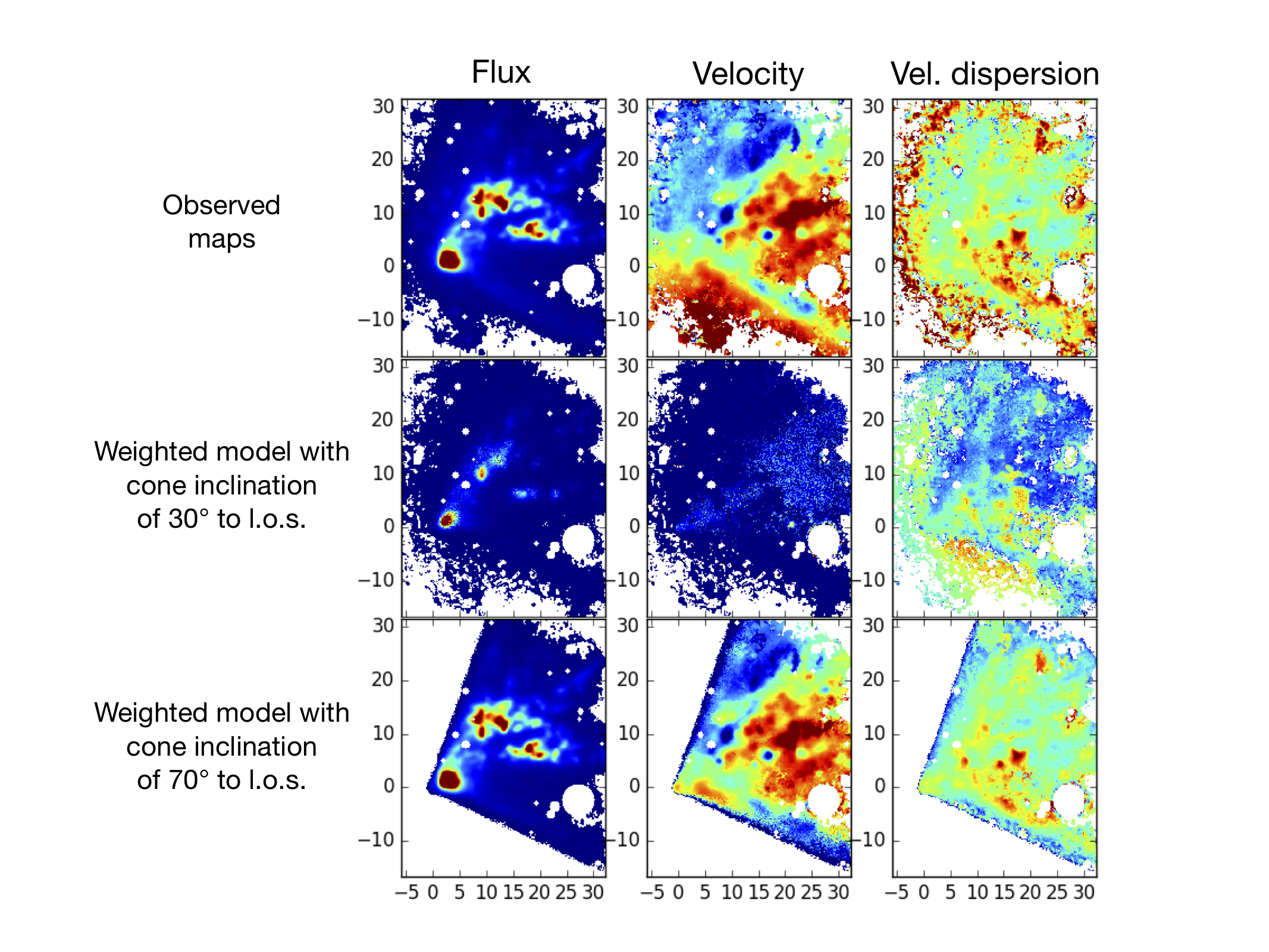} 
 \caption{Test of our new 3D outflow ``tomographic'' reconstruction on Circinus MUSE data. Maps of \oiii\ flux (left panels), velocity (middle; first-order moment of velocity of line profile) and velocity dispersion (right), respectively, from observations (top panels) and reconstructed from two models with constant velocity and hollow conical outflow geometry with inner and outer apertures of 40$\degree$ and 60$\degree$, respectively, one model having an inclination to line of sight (l.o.s.) of 30$\degree$ (middle panels), the other of 70$\degree$ (lower panels).}
   \label{fig:fig2}
\end{center}
\end{figure}

{\underline{\it New 3D kinematic model}}.
The extensive spatially-resolved information provided by MUSE on outflow kinematics allows us to obtain information on their geometry and structure. For instance, in \cite{Venturi:2017aa} the comparison between the observed MUSE maps and a very simple kinematic toy model suggested a hollow conical geometry for an outflow.
Only a few attempts of 3D-modelling of outflows from spatially resolved data, to take into account projection effects and infer the outflow kinematic structure, exist in  the  literature (e.g. \citealt{Crenshaw:2015aa}). These usually adopt a 
velocity field defined as a function of distance from the nucleus
and a smoothly distributed medium, convolve the model with instrument properties (PSF, bin on slit or spaxel size) and fit it  to spatially resolved data. 
However, the actual observed distribution of gas in outflows is not smooth but really clumpy and the velocity fields may be quite complex (e.g. Fig.~\ref{fig:fig2}, top panels). Such complex velocity fields may either arise from a real defined 3D velocity structure of the outflow or be an effect of the clumpy line-emission along the line of sight.

Here we propose a new approach to 3D-model outflows from spatially resolved observations, and  test it on MUSE data from our survey.
We consider a 3D outflow geometry and velocity field, uniformly populated with gas clouds, we transform it to the reference frame as seen by the observer, we bin it to the MUSE spaxel size and smooth to its seeing-limited resolution. At this point the uniform model is weighted directly on the observed cube of the emission-line profile in ($x$,$y$,$v$) space, according to the flux measured in each spaxel where a cloud is ``observed'', and sky-projected maps can be generated to compare with the observed ones. This procedure then allows a ``tomographic'' reconstruction of the outflow 3D structure following the assumption of a given velocity field.

To investigate possible degeneracies of this process, we  preliminary tested it on MUSE data of Circinus galaxy, and the results are shown in Fig.~\ref{fig:fig2}. They show that by choosing a wrong geometry and inclination for the outflow model, the sky-projected maps obtained from it (middle panels) do not reproduce well the observed ones (upper panels). By finding the appropriate model configuration, the observed maps can instead be very well reproduced (lower panels).
This first test on Circinus MUSE data shows that the complex velocity structures observed in the maps can be accounted for only by the effect of gas clumpiness on a radial and constant velocity field, without invoking complex kinematical effects.
We are currently further testing this approach for possible degeneracies by probing different combinations of geometries and velocity fields on different galactic outflows. 

\section{Conclusions}
We briefly reviewed the challenges in measuring outflow properties and their observational evidence from high to low redshift. Outflows are found to be ubiquitous in the ionised gas and multi-phase. The molecular and neutral atomic phases, when detected, dominate the mass of outflowing gas on galactic scales. While few indications of outflows damping star formation are found, clear observational evidence that they are capable of globally affecting star formation processes in the galaxy population are still missing. On one hand, reducing the uncertainties on measured physical properties and obtaining the total outflow mass budget from multi-phase studies is pivotal for a comparison with models and to better constrain the impact of outflows on host galaxies. On the other hand, the effect of outflows on star formation might be more subtle, for instance by having an effect on different timescales than those when outflows and star formation coexist (``delayed'' feedback) or by acting on longer timescales through a cumulative effect made up of multiple outflow and AGN activity episodes.

\bibliographystyle{aa_custom}

\end{document}